\documentclass{aa}
\usepackage{graphicx}
\usepackage{threeparttable}

\usepackage{txfonts}
\usepackage{xcolor}
\usepackage{soul}

\usepackage{float}
\usepackage[section]{placeins}
\usepackage{afterpage}

\usepackage{multicol}

\usepackage{tikz}

\usepackage[colorlinks=true, linktocpage, linkcolor={blue!60!black}, citecolor={blue!60!black}, urlcolor={blue!60!black}]{hyperref}

\newcommand\chandra{{\it Chandra\,}}
\newcommand\ergs{\ifmmode {\rm~erg\ s}^{-1} \else  ~erg s$^{-1}$\fi}
\newcommand\Zsun{\ifmmode Z_{\odot} \else $M_{\odot}$\fi}
\begin{document} 
\title{Deciphering the ultra-steep spectrum diffuse radio sources discovered in the cool-core cluster Abell~980}
\titlerunning{Ultra steep radio sources in Abell 980}
\authorrunning{Salunkhe et al.,}

   \author{Sameer Salunkhe\inst{1}, 
        Surajit Paul\inst{1}\fnmsep\thanks{Email:surajit@physics.unipune.ac.in},
        Gopal-Krishna\inst{2},
        Satish Sonkamble\inst{3}, and
        Shubham Bhagat\inst{4}}

     \institute{Department of Physics, Savitribai Phule Pune University, Pune 411007, India\\ \email{surajit@physics.unipune.ac.in; sameer@physics.unipune.ac.in}
     \and
            UM-DAE Centre for Excellence in Basic Sciences, University of Mumbai, Vidyanagari, Mumbai-400098, India
         \and
            INAF-Padova Astronomical Observatory, Vicolo dell’Osservatorio 5, I-35122 Padova, Italy 
             \and
             Thüringer Landessternwarte, Sternwarte 5, 07778 Tautenburg, Germany
            \\
             }

   \date{Received September 15, 1996; accepted March 16, 1997}

 
\abstract {
Clusters of galaxies are excellent laboratories for studying recurring nuclear activity in galactic nuclei since their hot gaseous medium can vastly prolong the detectability of their radio lobes via a better confinement. We report here a multi-band study of the sparsely studied galaxy cluster Abell 980, based on our analysis of {\it{Chandra}} X-ray and the GMRT (150 and 325 MHz) and EVLA (1.5 GHz) radio archival data, revealing an unusually rich phenomenology. It is shown to be a quasi-relaxed cluster with a cool core ($T\sim4.2$ keV) surrounded by a hot and extensive intracluster medium (ICM) at $T\sim6.8$ keV. The radio emission shows a rich diversity, having (i) two large diffuse sources of ultra-steep spectrum (USS), extending to opposite extremities of the ICM, each associated with an X-ray brightness discontinuity (cold front), (ii) a bright radio-double of size $\sim55$ kpc coinciding with the central BCG, and (iii) a diffuse radio source, likely a mini-halo of size $\sim110$~kpc around the BCG which possesses a huge ellipsoidal stellar halo of extent $\sim 80$~kpc. The association of cold fronts with two highly aged (~ 260 Myr) USS sources in a cool-core cluster, makes it a very rare system. These USS sources are probably radio lobes from a previous episode of jet activity in the BCG, driven buoyantly towards the outskirts of the X-ray halo, thereby creating the cold fronts. A deeper radio image of this cluster may provide a rare opportunity to verify the recently proposed alternative model which explains radio mini-haloes as the aggregate radio emission from Type I supernovae occurring in the giant stellar halo extended across the cluster core.
}


\maketitle
%

\section{Introduction}\label{intro}

Clusters of galaxies are natural habitat of diffuse steep-spectrum radio sources ($\alpha < -1.2$), usually identified as the quasi-confined relic lobes of once active radio galaxies \citep{1973MNRAS_baldwin, Murgia_2011A&A, 2017MNRAS_Godfrey}. Detecting these sources, especially their ultra-steep spectrum (USS) subset, requires (low-frequency) radio imaging with high sensitivity and dynamic range at a fairly high spatial resolution. One signpost of such relic lobes is the `cavities' sometimes detected in the X-ray images of galaxy clusters and groups (e.g. \citealt{2007ARA&A_McNamara}). Their radio detection is facilitated due to the brightness enhancement resulting from adiabatic compression by a passing shock wave usually launched during the merger(s) of the host cluster with another galaxy group/cluster \citep{Ensslin_2001A&A}. Such `radio phoenixes' are thus not expected to be found in `cool core' clusters.  On the other hand, such clusters sometimes exhibit radio `mini-haloes', the diffuse radio sources of steep spectrum extending to $\sim100$~kpc around the Brightest Cluster Galaxy (BCG) (e.g. \citealt{Giacintucci_2019ApJ}). While, studies of USS  radio sources in clusters have now got a huge fillip, with the availability of dedicated low-frequency radio telescopes, such as LOw Frequency ARray (LOFAR) and the upgraded Giant Metrewave Radio Telescope (uGMRT) (e.g. \citealt{Savini_2018MNRAS}), archival research continues to surprise us by revealing extraordinary objects and here we report one such finding in a comparatively low-mass cluster Abell 980.

Abell 980 (A980; RXC~J1022.5+5006) at $z$ = 0.1582 \citep{1991Lebedev}, has an SZ mass $M_{500}^{SZ} = 4.73_{-0.32}^{+0.29} 
\times 10^{14} \rm{M_{\odot}}$ \citep{palnck2016A&A}, a high X-ray luminosity  ($L_X=7.1\times10^{44}$~erg~s$^{-1}$) and an average ICM temperature (7.1 keV; \citealt{Ebeling_MNRAS_1996}). By processing the WENSS survey data, with a resolution of $54\arcsec$ at 330~MHz, \citet{2009ApJRudnick} showed that A980, has a luminosity $P_{\rm{330MHz}}=5.6 \pm 1.3 \times 10^{24}$~W Hz$^{-1}$ at 330 MHz, and some marginally resolved faint emission underlies a dominant radio peak. 

This paper reports the discovery of a rich assembly of steep to ultra-steep spectrum radio sources in A980, by analysing the archival data from GMRT  (150 \& 325~MHz), EVLA (1.5 GHz) and {\it Chandra} (soft X-rays). \S~\ref{sec:data} describes the data analysis, results of which are summarized in \S~\ref{sec:res} and discussed in \S~\ref{Discussion}, followed by our main conclusions given in \S~\ref{conc}. A $\Lambda$CDM cosmology is assumed with parameters $H_{0}$=$70~{\rm{kms^{-1} Mpc^{-1}}}$, $\Omega_M$=$0.3$, $\Omega_{\Lambda}$=$0.7$. The physical scale for the images is 2.73~kpc/arcsec.
 
\vspace{-0.2cm}

\section{Observations and data reduction}\label{sec:data}

\subsection{The analysis of the radio data}\label{sec:radio_anal}

We have analysed a total of 100 minutes (150 MHz) and 314 minutes (325 MHz) of on-source GMRT archival data (Project: ddtB020~\&~17\_073) from RR and LL correlations using the Source Peeling and Atmospheric Modeling (SPAM) pipeline (for details, see \citealt{2017Intema_SPAM}). This pipeline takes care of direction-dependent variations in visibility amplitude and phase across the field of view, due to the antenna beam pattern and the ionosphere. In the initial SPAM steps, the calibrator 3C48 was used for flux density and bandshape calibration. The available 56 minutes of  EVLA L-band (1-2~GHz, B-array) data (Project: 15A-270 and 12A-019) from RR and LL correlations were imaged using the Common Astronomy Software Applications (CASA) calibration pipeline and performing several rounds of phase-only self-calibration and one amplitude and phase self-calibration. Images with multiple angular resolutions were made using the various weighting and uv-taper parameters listed in Table~\ref{tab:imaging}. The spectral index and error maps were produced using the IMMATH task of CASA following the standard relations and taking the data from different frequencies (Appendix~\ref{apdx:radio-spectrum}). 

Since the radio emission from A980 is found to consist of a diffuse source coinciding with the BCG and several discrete sources in the vicinity, an enhanced diffuse radio emission map was also generated, by filtering the discrete source contributions. We broadly followed the procedure given in \citet{2014van_Weeren_Phoenix}. Firstly, an image of the discrete sources was made by only using uv spacing $>3\:\rm{k\lambda}$. The corresponding visibilities were then subtracted from the observed UV data and the final low-resolution image was made setting the `Briggs' weighting, robust = 0 and a uv-taper of $10\:\rm{k\lambda}$.

All measurements on the images were carried out using the CASA viewer. The flux densities (S) were measured within the $3\sigma$ isophotes of a given source and the errors ($\sigma_S$) were calculated as $\sigma_S=\sqrt{(0.1 S)^2 + N\sigma^2+ \sigma_{sub}^2}$, where N is the number of beams across the diffuse emission, $0.1S$ is assumed to be the error due to calibration uncertainties and $\sigma_{sub}$ is the uncertainty arising from the compact source subtraction procedure (applied to the point source subtracted maps only).

\begin{table} 
\caption{Parameters of the radio images} %
\label{tab:imaging}      
\centering          
\begin{tabular}{lcccr}     
\hline\hline       
Frequency &  Robust & uv-taper & FWHM & rms \\
(MHz)& (briggs) & ($\rm{k\lambda}$) & (in \arcsec; PA in $^\circ$)& ($\mu$Jy/beam)\\
\hline  
150  &  -0.5 &  8 & $33\times21$; 59  & 5500\\
325 & uniform & -- & $6.7\times5.7$; 5$^h$ & 140\\  
  &  -2 &  -- & $8.2\times7.3$; 7$^m$ & 100\\
  &  0 & 10  & $19\times16$; 61$^l$& 215\\
1500  & -2 & -- & $2.8\times2.5$; 63$^h$ & 30\\
  &  0 & 10 & $9.0\times8.7$; 31$^l$ & 25\\
\hline
\hline
\vspace{-0.3cm}
\end{tabular}
\begin{tablenotes}
      \small
\item \textbf{Note}: Resolutions: h- high, m- medium, l- low 
    \end{tablenotes}
\end{table}

\subsection{Analysis of the X-ray data}\label{apdx:x-ray-anal}
The level-1 event file of A980 from {\it Chandra} data archive (14\,ks; ObsID 15105) was reprocessed following the standard data-reduction routine of CIAO\footnote{\color{blue}{{http://cxc.harvard.edu/ciao}}}~4.11 and employing the latest calibration files CALDB 4.8.3. Events were screened for cosmic rays using the ASCA grades, and were reprocessed by applying the most up-to-date corrections for the time-dependent gain change, charge transfer inefficiency, and degraded ACIS detector quantum efficiency. Periods of high background flares exceeding 20\% of the mean background count rate were identified and removed using the {\tt lc\_sigma\_clip} algorithm. This yielded 13.5\,ks of net exposure time. The standard blank-sky data sets\footnote{\color{blue} {http://cxc.harvard.edu/ciao/threads/acisbackground/}} were processed and re-projected to the corresponding sky positions and normalized to match the count rate in $10-12$~keV energy range. Point sources across the ACIS field, except the nucleus of A980,  were identified using the CIAO tool {\tt wavdetect} and removed \citep[for detail see][]{2015Ap&SS.359...61S}. Thereafter, exposure correction was applied on the cleaned X-ray image (i.e. free from point sources and flares), using the mono-energetic exposure map created at 1~keV.

\vspace{-0.1cm}
\section{Results}\label{sec:res}

\subsection{The radio properties}\label{sec:radio}
\subsubsection{Radio structural properties of A980} \label{sec:radio_properties}

\begin{figure}

\includegraphics[width=0.49\textwidth]{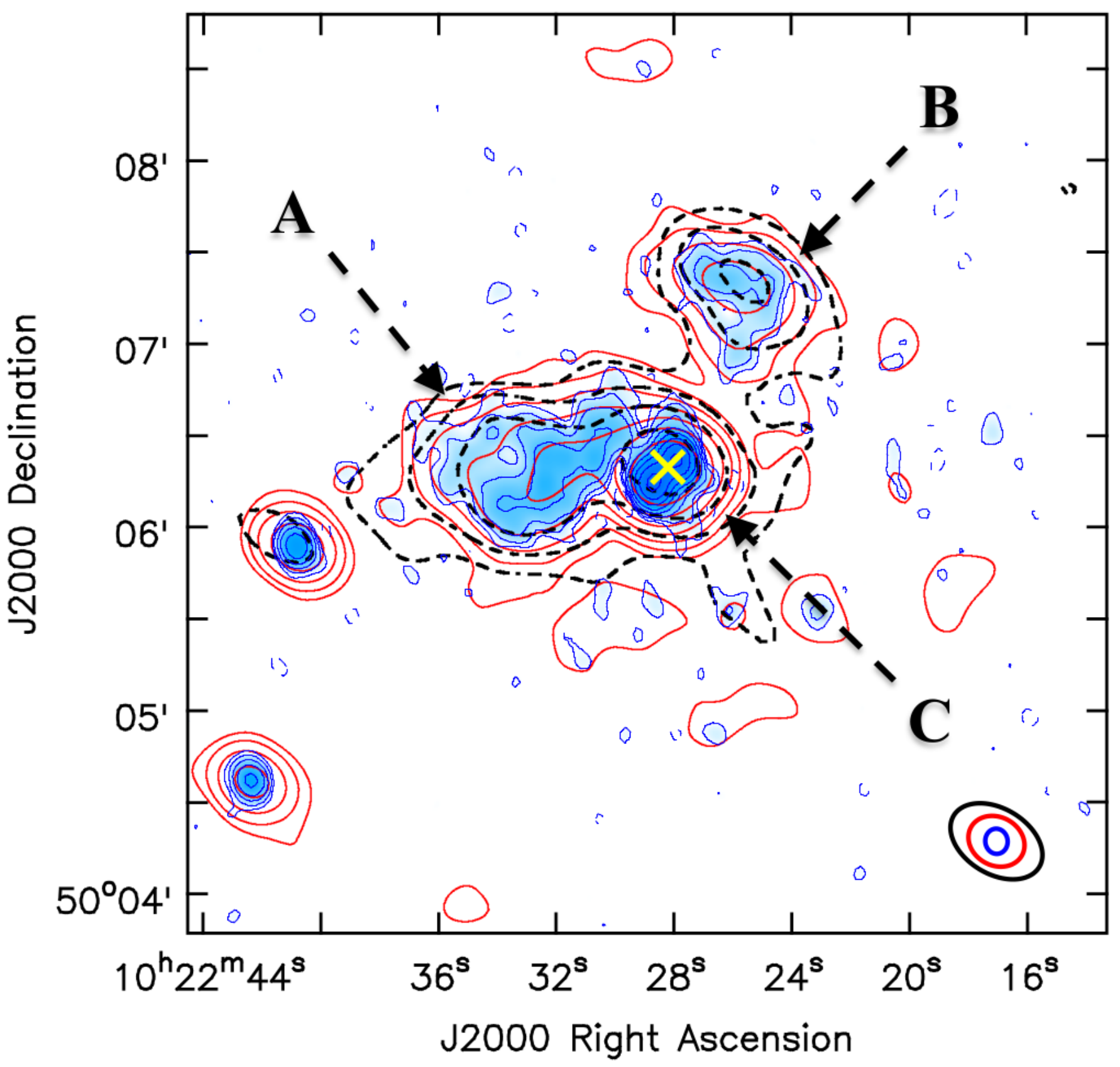}
\caption{The GMRT 325~MHz medium-resolution image (blue colour and contours drawn at $\pm3, 6, 12, ... \times \sigma$; $\sigma = 100~\mu$Jy/beam) is overploted with the 325~MHz low-resolution map (red contours) at $\pm3, 6, 12, ... \times \sigma$ ($\sigma = 215~\mu$Jy/beam). The 150~MHz GMRT map is shown with black dashed contours drawn at $\pm3, 6, 12,... \times \sigma$ ($\sigma = 5.5$~mJy/beam). 
}\label{fig:GMRT_high_low}
\vspace{-0.1cm}
\end{figure}

From Fig.~\ref{fig:GMRT_high_low}, the radio emission is concentrated in the regions, A, B and C which encompasses the BCG (yellow cross). These sources are seen at all the 3 radio frequencies, except B which is undetected at 1.5 GHz (see below). 

\begin{figure}
\includegraphics[width=0.49\textwidth]{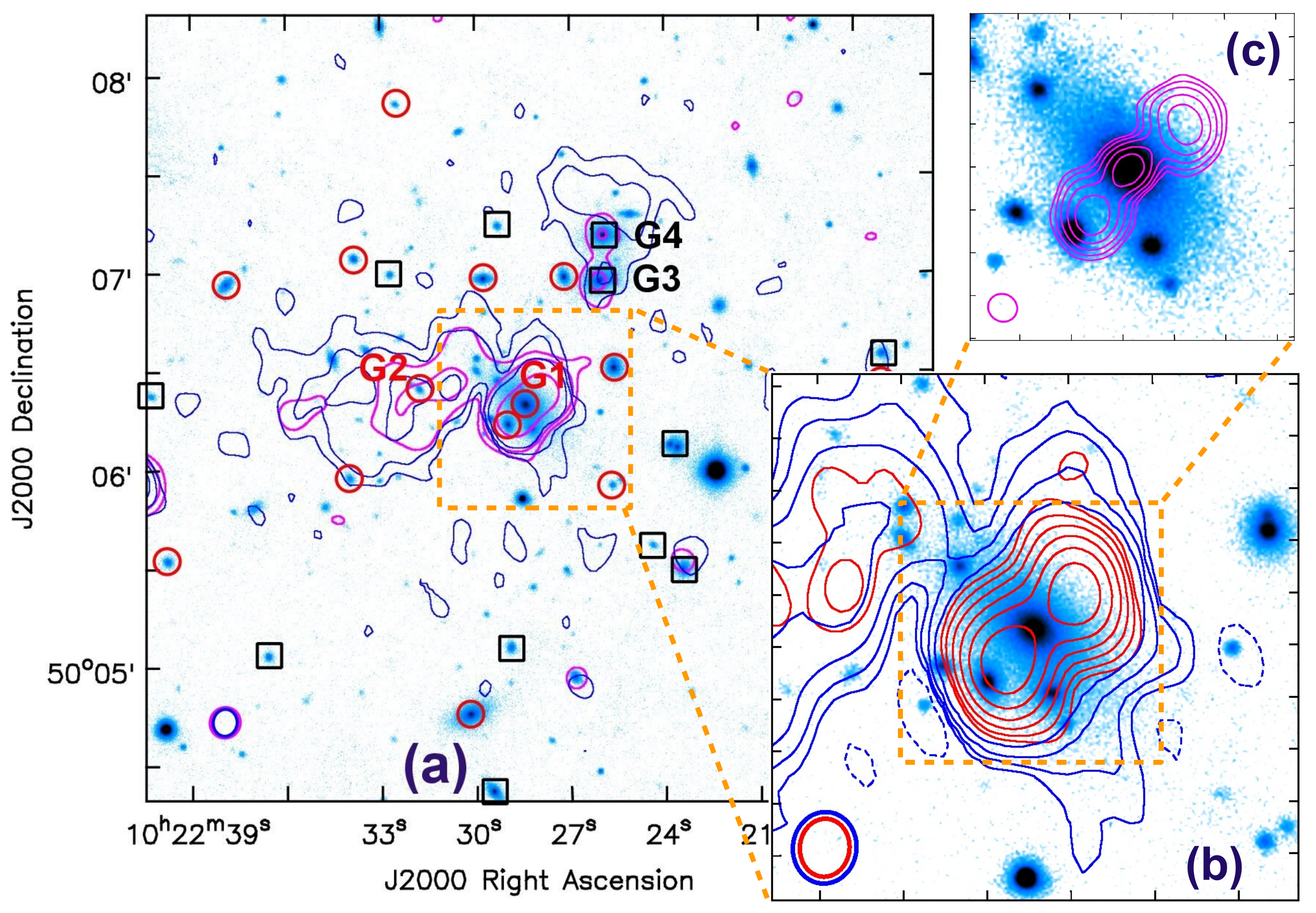}
\caption{\textbf{Panel (a):} Pan-STARRS `i' band optical image is superimposed with the low-resolution VLA L-band map (magenta contours at $ 3, 9, 81 \times(\sigma=25~\mu$Jy/beam)) and the 325 MHz medium-resolution GMRT map (blue contours at $ 3, 9, 27 \times(\sigma=100~\mu$Jy/beam)). The cluster galaxies  ($z=0.1582\pm0.0035$; red circles) and foreground/background galaxies ($z<0.1547$ or $z>0.1617$; black squares) are shown within the field. The estimated redshift range ($\Delta z = 0.0035$) equivalent to the velocity dispersion ($v_{rms}=1033$~km/s) of the cluster \citep{2013ApJRines}.  \textbf{Panel~(b):} Contours (red) of the GMRT 325~MHz map with uniform weighting, plotted at $5,10,20,... \times(\sigma=140~\mu$Jy/beam) and those with robust -2 setting (blue), plotted at $3,6,12,... \times(\sigma=100~\mu$Jy/beam). \textbf{Panel~(c):} High-resolution VLA L-band contours are plotted at $3,6,12,...\times(\sigma=30~\mu$Jy/beam).}\label{fig:radio_high_opt}
\end{figure}

Using the medium-resolution GMRT map at 325 MHz (see Fig.~\ref{fig:GMRT_high_low}), the sizes of `A' and `B' were measured to be $190\times170$~kpc and $140\times140$~kpc, respectively. The 1.5~GHz VLA map (Fig.~\ref{fig:radio_high_opt}) detects only the emission associated with the BCG and the brighter parts of `A'. Unlike B, the sources A and C are not clearly separated in the GMRT 150 MHz map but their flux densities can still be measured fairly accurately and these are $261\pm26$ mJy and $117\pm12$ mJy, respectively. From Fig. \ref{fig:radio_high_opt}, the region C at 325 MHz and 1.5 GHz is comprised of (i) a `double' radio source of size $\sim20$~arcsec at position angle PA=135 deg, straddling a core which is seen only at 1.5 GHz and which coincides with the BCG, and (ii) an underlying faint and hence not so well imaged mini-halo candidate measuring 0.9 arcmin (north-south) (Figures \ref{fig:radio_high_opt}b~\&~\ref{fig:A980_temerature}; Table~\ref{tab:diffuse_emission}).

\vspace{-0.2cm}
\begin{table} 
\caption{Measured parameters of the main radio components} %
\centering    
\small
\begin{tabular}{p{0.6cm}ccccc}
\hline       
Source &  \multicolumn{3}{c}{Flux density (mJy)} & \multicolumn{2}{c}{Spectral index}\\
& [150MHz & 325 MHz & 1.5 GHz] & [$\alpha_{\rm{150}}^{\rm{325}}$ & $\alpha_{\rm{325}}^{\rm{1500}}$]\\
\hline  
A &  $261\pm26$ & $54.3\pm5.5$ & $1.4 \pm 0.2$ & $-2.0 \pm 0.3$ &$-2.4 \pm 0.2$ \\
B & $117\pm12$ & $20.4\pm2.1$ & <  $0.89$ & $-2.3 \pm 0.3$  & $< -2.1$  \\
C &  $281\pm28$ & $120\pm12$ & $13.8\pm1.4$ & $-1.1\pm0.3$ & $-1.4\pm0.1$\\
\hline
\hline
\end{tabular}
\label{tab:diffuse_emission} 
\end{table}

\subsubsection{Radio spectral properties}\label{radio_spectral_properties}

\begin{figure}
\includegraphics[width=0.48\textwidth]{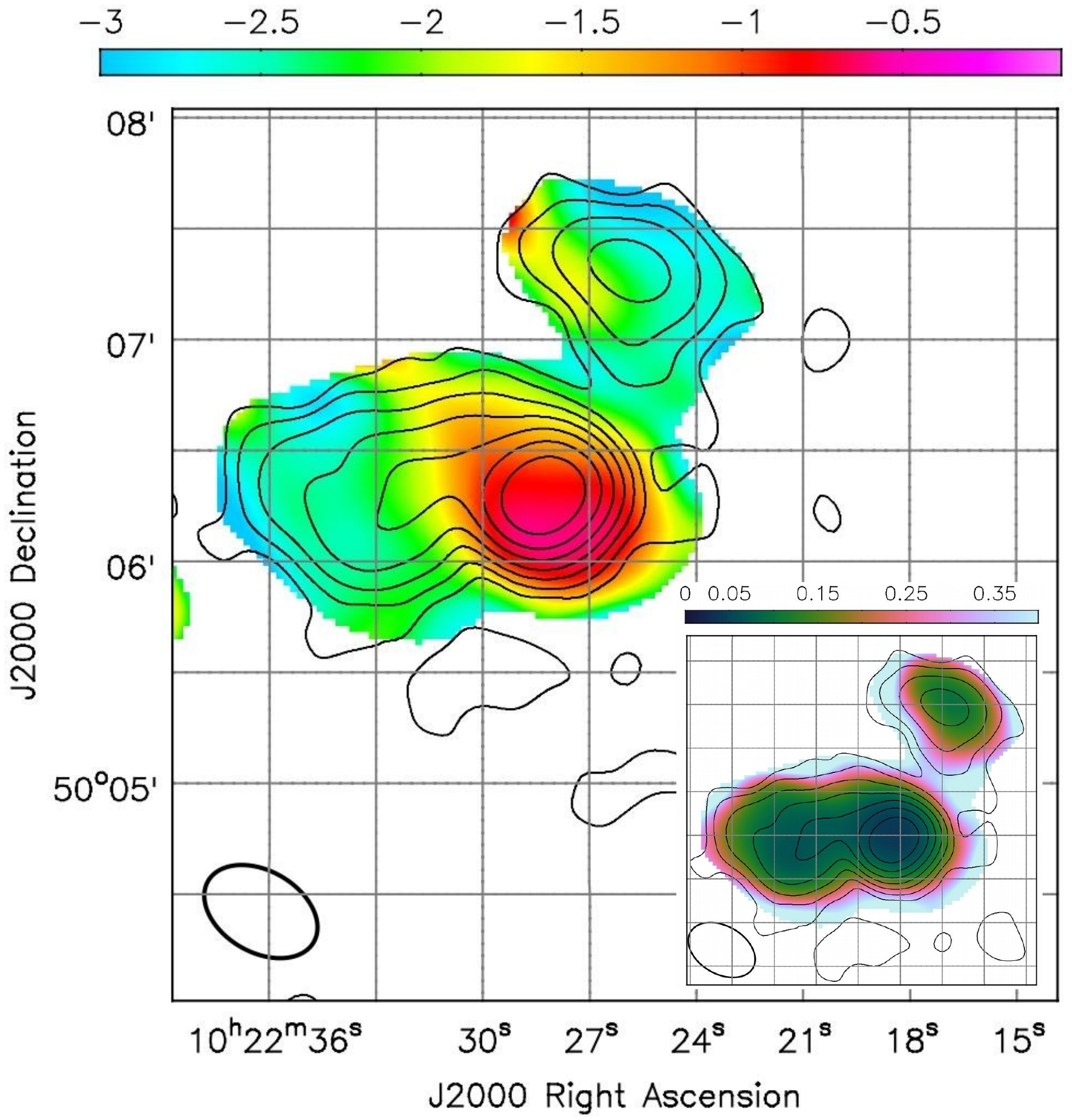}

\caption{GMRT spectral index and error (inset) maps (150-325 MHz), with beam size of $33\arcsec\times22\arcsec$, PA$=60^{\circ}$. The contour levels are same as in the low-resolution map at 325 MHz (Fig. \ref{fig:GMRT_high_low}).}\label{fig:spix-TGSS-GMRT}
\end{figure}

For the entire complex seen in Fig. \ref{fig:GMRT_high_low}, spectral index was computed, firstly, by combining the GMRT 150 MHz and 325 MHz (low-resolution) maps, integrating within 3$\sigma$ contours, yielding $\alpha^{325}_{150}= -1.56 \pm 0.25$. Secondly, we combined the medium-resolution GMRT 325 MHz map with the low-resolution VLA map at 1.5 GHz (see Table~\ref{tab:imaging}), restricting the integration to within their overlapping portions, which gave
$\alpha^{325}_{1500}= - 1.49 \pm 0.13$. Spectral indices of the individual sources are given in Table~\ref{tab:diffuse_emission}, one of which is an upper limit, owing to the non-detection of B at 1.5 GHz
(Fig. \ref{fig:radio_high_opt}a).
The spectra of both A and B remain very steep even at low frequencies, with $\alpha^{325}_{150}= - 2.0 \pm 0.3$ for A, and {$- 2.3 \pm 0.3$} for B. This is highly reminiscent of relic radio sources (e.g. \citealt{Murgia_2011A&A}). Fig. \ref{fig:spix-TGSS-GMRT}, displays the variation of $\alpha^{325}_{150}$ across the sources A and B, starting from $\approx -1.6$ near the inner region adjacent to C, it steepens to $\approx -2.7$ near their extremities.
For both these USS sources, we have estimated spectral ages since the cessation of energy input ($\sim265\pm33$~Myr for `A' and $\sim260\pm45$~Myr for B), following the procedure followed
in \citet{2004AA_jamrozy} and adopting the JP model (details in Appendix~\ref{apdx: spectral_age}). The derived ages are lower limits since the spectral break frequency was taken to be 150 MHz, the lowest observed frequency. The derived equipartition magnetic field for A and B are $2.6~\mu$G and $3.2~\mu$G, respectively. The small double source hosted by the BCG also shows a rather steep spectrum ($\alpha^{325}_{150}\gtrsim-1.1 $ and $\alpha^{1500}_{325}=-1.3 \pm 0.1$)
and its estimated spectral age is $\sim 70$ Myr.

\subsection{The X-ray properties}\label{xray_properties}

\begin{figure*}
\includegraphics[width=0.4\textwidth]{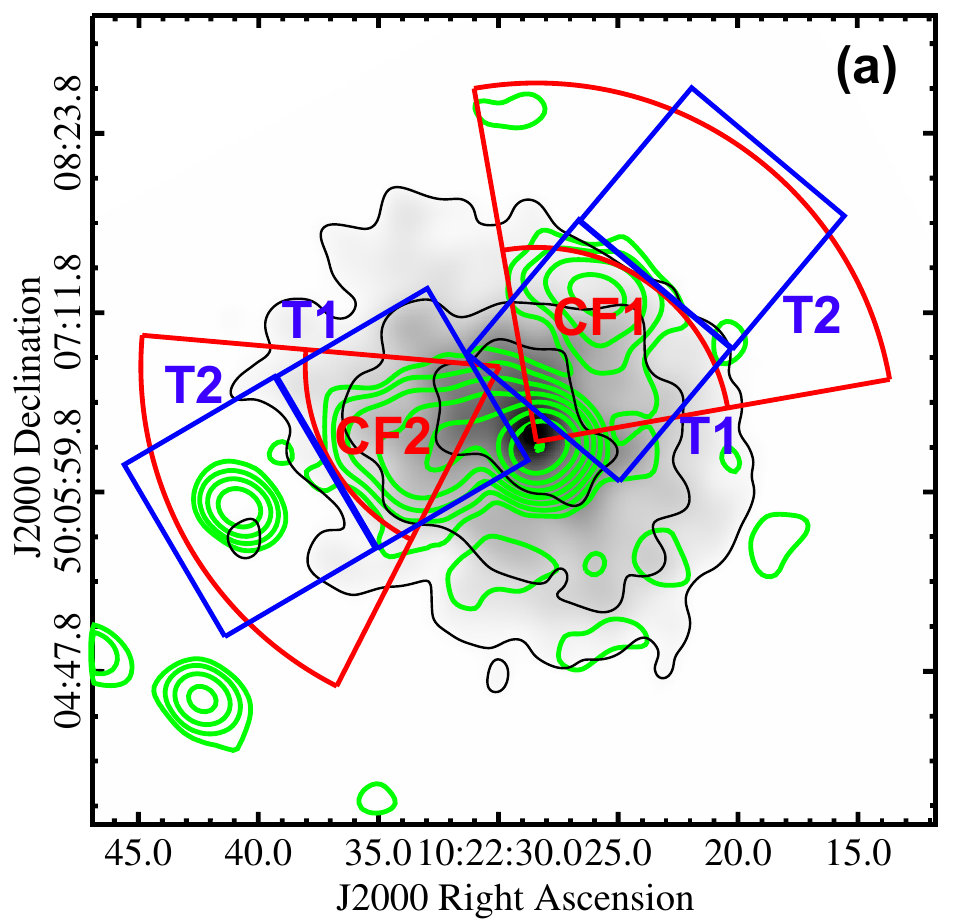}\hspace{-0.15cm}
\includegraphics[width=0.32\textwidth]{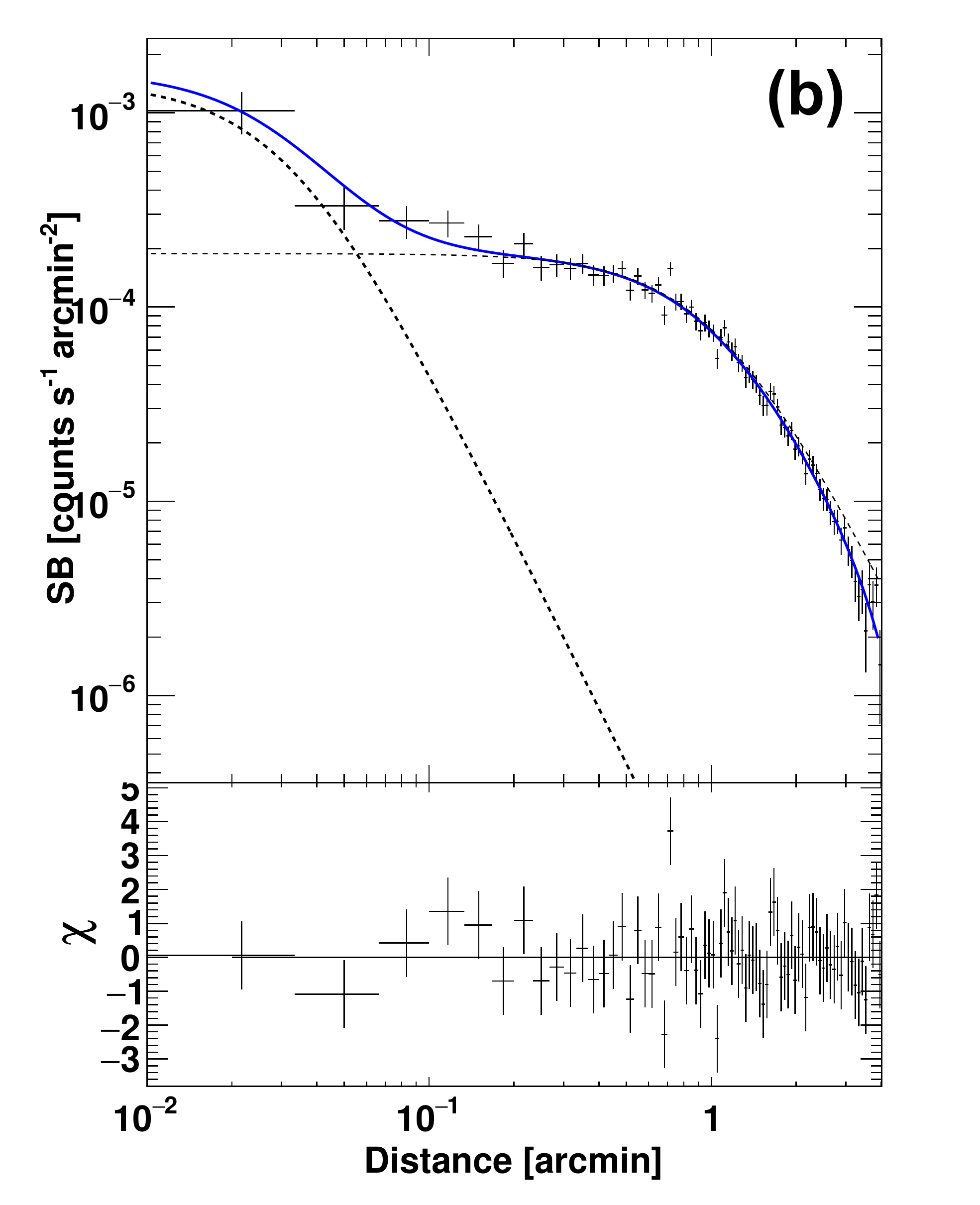}\hspace{-0.4cm}
\includegraphics[width=0.32\textwidth]{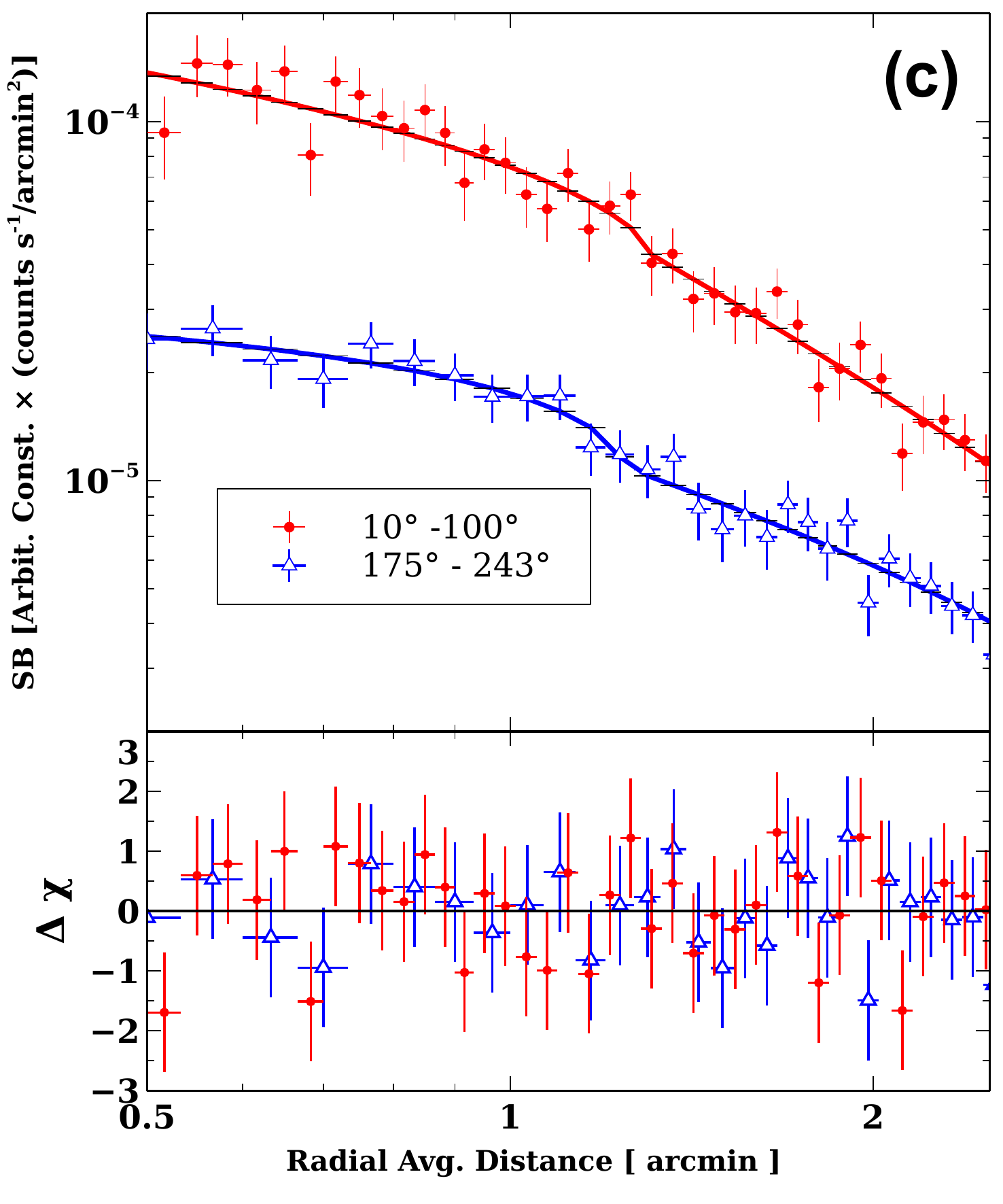}

\caption{{\bf Panel~(a):} The {\it Chandra} X-ray map in gray colour and black contours, over-plotted with the GMRT 325~MHz (low-resolution) contours (green). The sectors (10\degr-100\degr (CF1) and 175\degr-243\degr (CF2)) along which the source `B' and `A' are extended, respectively, are marked as red cones. Blue boxes denote the pre (${\rm T_{2}}$) and post (${\rm T_{1}}$) cold front regions (see text). {\bf Panel~(b):} Shows the azimuthally averaged X-ray surface brightness (SB) profile, with the best-fit double $\beta$-model. The individual $\beta$-models are shown with dotted lines and the residuals are shown in the bottom panel. {\bf Panel~(c):} The SB profiles for the two conal sectors and the fitted broken power-law models are shown with red and blue curves, respectively.}\label{fig:xray-fig}
\end{figure*}

From the \textit{Chandra} archives, we extracted the $0.5-8$\,keV spectrum of the region within 150 arcsec, (408\,kpc) of the X-ray peak, after excising the central 2 arcsec region at the peak associated with the nucleus of the double radio source within C, as well as other obvious point sources (details in Appendix~\ref{apdx:glob-xray}). Figure~\ref{fig:xray-fig}a presents the 5$\sigma$ Gaussian-smoothed \chandra image ($0.5-3.0$\,keV), revealing a roughly circular extended source. However, a few faint substructures and a mild elongation can be noticed towards north-east of the X-ray peak (see Fig.~\ref{fig:xray-fig}a). For the ICM, we estimate the best-fit global elemental abundance of $0.33\pm0.08$ $\Zsun$, a temperature (T) of $6.8\pm0.4$~keV and an integrated $0.5-8$\,keV luminosity of $5.9\pm0.1\times10^{44}$\ergs (Appendix \ref{apdx:glob-xray}). The radial surface-brightness profile, assuming a spherically symmetric ICM (see Fig.~\ref{fig:xray-fig}b; and Appendix~\ref{apdx:xray_SB} for details), is well represented by a double $\beta$-model, showing a prominent `core' of high surface brightness. The best-fit model parameters are given in Table~\ref{tab:jump_SB}.

\subsubsection{The X-ray edges} \label{xray_edge}

\begin{table*}
\caption{Properties of the cold fronts seen at A and B (details in  Table~\ref{tab:jump_density})}
\centering
\small
\begin{tabular}{cccccccccc}\hline
Conal &Radial& Density enhancement & $T_1$ & $T_2$* & Temperature Jump & $P_1$ & $P_2$ \\
sector& position & factor & (keV) & (keV)& ($\Delta T=T_2$-$T_1$ (keV)) & $10^{-11}$ erg cm$^{-3}$ & $10^{-11}$ erg cm$^{-3}$  \\ \hline
10\degr - 100\degr (A) & 1.30\arcmin & $1.20 \pm 0.12$ &$5.77 \pm 0.48$ & $10.90\pm 3.33$ & $5.13 \pm 3.36$ & $4.49 \pm 0.40$ & $5.38 \pm 0.72$ \\ 
$\chi^2$/dof & -- & -- &$56.33/57$ & $23.82/24$ & -- & -- & -- \\ 
175\degr - 243\degr (B) & 1.32\arcmin &  $1.35 \pm 0.15$ & $5.53 \pm 0.52$ & $12.22 \pm 3.87$ & $6.69 \pm 3.90$ & $1.04 \pm 0.10$ & $1.03\pm0.32$ \\ 
$\chi^2$/dof & -- & -- &$58.77/57$ & $21.74/25$ & -- & -- & -- \\
\hline
\end{tabular}\\
* - metallicity fixed to 0.33 $\Zsun$ i.e average metallicity of this cluster. 
\label{tab:jump_temperature}
\end{table*}

As seen in Fig.~\ref{fig:xray-fig}a, the outer edges of the diffuse radio sources A and B  (marked by the conal sectors, drawn in red colour) appear roughly aligned with the X-ray contours near the ICM outskirts. In the radial X-ray surface-brightness profiles for these conal sectors encompassing the radio sources A and B, we have detected discontinuities (Fig.~\ref{fig:xray-fig}c). To examine any change in conditions across these X-ray `edges', we defined a pair of square-shaped regions (blue boxes) on opposite sides of each edge (Fig. \ref{fig:xray-fig}a) and determined the electron density, temperature and pressure within these boxes (see Appendix~\ref{apdx:xray_edges} for the methodology). The de-projected electron density profile along the sectors CF1 and CF2, as well as the temperature determined for the pairs of boxes, show significant discontinuity in both parameters. This indicates physical differences in the ICM on the two sides of the outer edge of both A and B (Fig.~\ref{fig:xray-fig}c and Table~\ref{tab:jump_temperature}). In both cases, the box on the inner side of the radio edge has a lower temperature (T1), whereas no significant change is seen in pressure (see Appendix~\ref{apdx:xray_edges} and Table~\ref{tab:jump_temperature}). This is strongly suggestive of a `cold front' for each of these two X-ray discontinuities. Note also a marginally significant radial increase in metallicity ($0.59 \pm 0.45$) found across the cold front CF1.

\subsubsection{The X-ray temperature map}\label{sec:xray_ktmap}
A 2D temperature map has been derived using the  contour-binning algorithm by \cite{sanders2006}. This algorithm generates a set of regions following the distribution of the surface brightness so that each region has nearly the same signal-to-noise ratio. A total of 9 regions were thus delineated, each with a signal-to-noise ratio of 20 ($\sim400$ counts per bin) from the $0.5-3.0$~keV image. We then extracted the X-ray spectrum for these regions and fitted with the model {\tt TBABS}$\times${\tt APEC} using the method described in Appendix~\ref{apdx:glob-xray}. While fitting, we fixed the metallicity 0.33 $\Zsun$ being the average metallicity of this cluster. The final map is shown in Fig.~\ref{fig:A980_temerature} overlaid with temperature values and corresponding uncertainties.

The temperature map reveals large spatial variations in T, indicating the presence of substructures. Although, the core is distinctly cool with a temperature $T=4.2\pm0.6$~keV (size $\sim100$~kpc), patches with temperature as high as $T\sim12$~keV are seen immediately north-east of the cool-core, roughly co-spatial with the few X-ray sub-structures mentioned in \S~\ref{xray_properties}. Overall, the cluster ICM has a high average temperature of $T\sim6.8$~keV as found in the present analysis (\S \ref{xray_properties}).


\section{Discussion} \label{Discussion}

From \S~\ref{sec:radio_properties}, the radio emission in A980 can be segregated into four main components: two large diffuse USS radio sources A and B, both having $\alpha\lesssim-2$, and a region C which consists of a smaller double radio source ($\sim55$~kpc) at the BCG and an underlying faint extended steep-spectrum radio source ($\sim100$~kpc). Below we outline some likely scenarios for their origin.

\subsection{Episodic AGN activity and the relic radio lobes} \label{Dying_radio_jets} 

From Fig.~\ref{fig:GMRT_high_low}~and~\ref{fig:radio_high_opt}, it can be seen that the USS sources A and B are diffuse and moderately luminous ($\sim3.7 \times 10^{24}$ W Hz$^{-1}$ and $\sim1.4 \times 10^{24}$ W Hz$^{-1}$ at 325 MHz, respectively), strongly reminiscent of relic lobes of radio galaxies that are no longer active (e.g. \citealt{2008MNRAS_jamrozy, Murgia_2011A&A, 2017MNRAS_Godfrey}). Galaxies G3 and G4 are seen close to B, however their redshifts ($z$ = 0.149556 and 0.152417) are significantly discrepant from that of A980. Furthermore, each of them has its own barely resolved radio counterpart seen in the VLA L-band low-resolution map at 1.5 GHz (see Fig.~\ref{fig:radio_high_opt}a), with the spectrum ($\alpha\lesssim-1$; see Appendix~\ref{apdx:radio-spectrum}) of an active point-like radio source. Therefore both these normal-spectrum radio sources are highly unlikely to be physically associated with the diffuse USS source B. The galaxy G2, which is centrally located within the USS source A is undetected in the VLA-FIRST survey. Moreover, its apparent magnitude $m_{r}=20.06\pm0.05$ (PANSTRASS-DR1, \citealt{chambers_2016arXiv}) corresponds to an absolute magnitude $M_r=-19.39\pm0.05$, at its redshift z=0.161662. This is 3.4 magnitudes under-luminous compared to the average absolute magnitude ($-22.8\pm0.5$) of the ellipticals hosting radio galaxies at $z < 0.5$ \citep{Treves_2005ApJ}, which makes it a very unlikely host of the source `A'. Thus, there are no viable optical identifications for A and B as independent radio sources. This, together with the similar spectral ages of 265 Myr (A) and 260 Myr (B) (\S \ref{radio_spectral_properties}), as well as the conspicuous radio extension of A towards B strongly suggest that the two USS sources are aged lobes of a common host galaxy. 

Considering the absence of likely host galaxies for A and B and given that intermittent nuclear activity is a common feature of radio galaxies (e.g.  \citealt{2017MNRAS_Godfrey, 2018MNRAS_turner}), it seems quite plausible that A and B are aged remnants of a powerful double radio source created by the BCG (G1 in Fig. \ref{fig:radio_high_opt}a) in its previous active phase, and both have risen to the ICM outskirts due to buoyancy pressure, while remaining confined due to the thermal pressure of the hot ICM. Concurrently, the galaxy has drifted towards the cluster centre where it has entered a new active phase resulting in the small double source with an associated compact radio core (Fig. \ref{fig:radio_high_opt}b \& \ref{fig:radio_high_opt}c). In absence of a clear tracer of a jet axis in the available maps, we considered the line joining the outermost parts of the two old lobes as an approximate location of the radio axis from where the BCG (G1) would have to move a distance of about 100 kpc to arrive at its present location near the cluster centre. This postulated trajectory of the drift is the natural radial direction for the infall towards the cluster centre as inferred from the X-ray peak which has a very small offset ($\sim2$ arcsec = 5.5 kpc) from the present location of the BCG. Note that in order to cover the (projected) distance of $\sim100$~kpc within the age of the two USS sources A and B created by it, the required drift velocity is
$\sim400$~km/s which is not exceptional for BCGs (e.g. \citealt{sohn_2020ApJ}) in quasi-relaxed clusters and certainly well within the observed velocity dispersions (1033 km/s) of the galaxies in this cluster \citep{2013ApJRines}.

\begin{figure}
\includegraphics[width=0.48\textwidth]{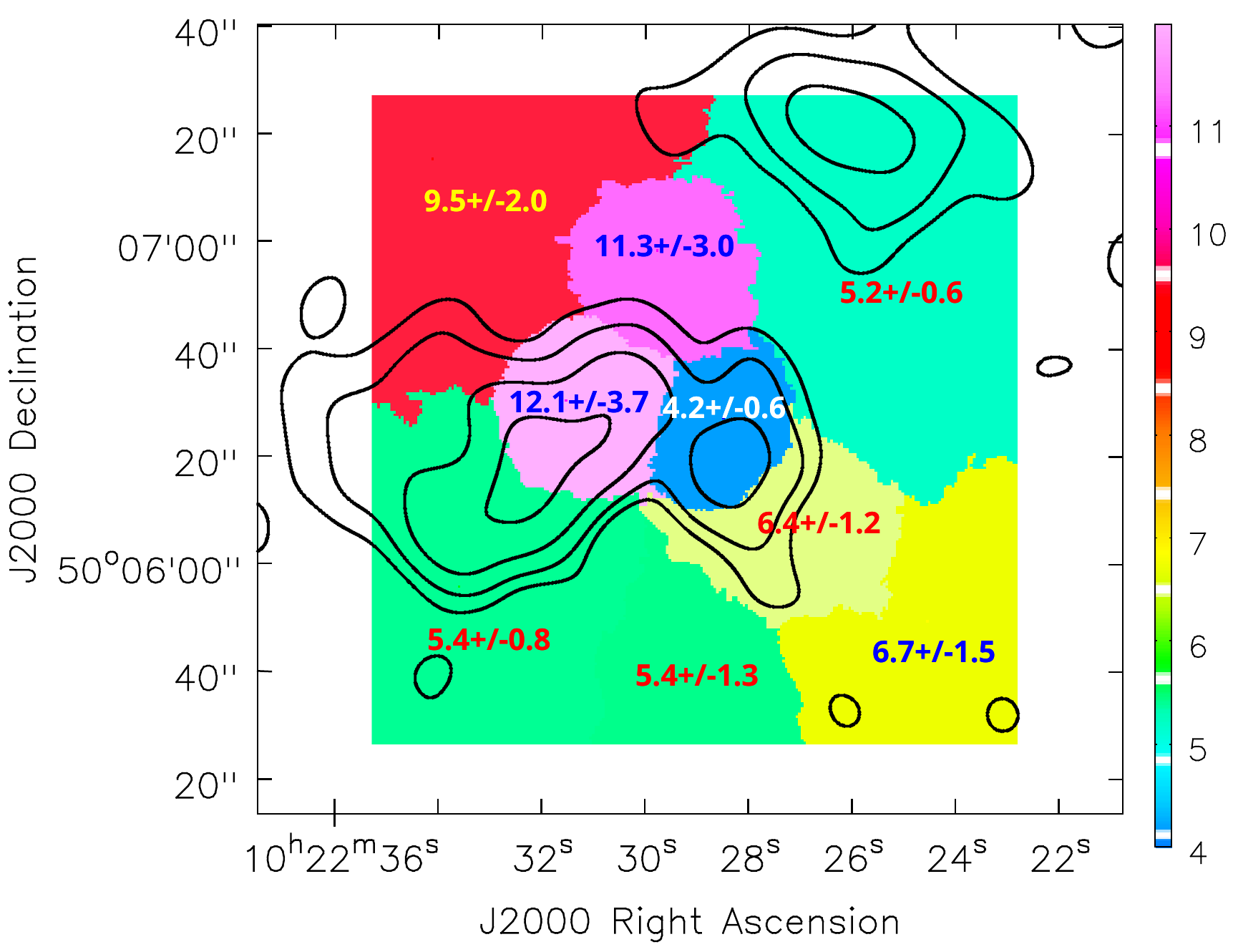}
\caption{{\it Chandra} X-ray temperature (keV) map overlaid with the radio contours (at $3,6,12,... \times 250~\mu$Jy/beam) of the point-source-subtracted 325~MHz GMRT map.}\label{fig:A980_temerature}\vspace{-0.3cm}
\end{figure}

\subsection{Fossil radio lobes and the X-ray edges in the ICM}

Discrete USS sources inside cluster ICM can also be detectable when fossil radio lobes remaining from a past episode of jet-activity in a cluster galaxy are subject to adiabatic compression caused by a passing shock-wave from the cluster's merger with another galaxy cluster or group \citep{Ensslin_2001A&A}. However, there is no evidence of a shock within the ICM of this cool-core cluster, particularly in the vicinity of the two USS sources (Fig. \ref{fig:xray-fig}). Nonetheless, other imprints of the two USS radio lobes are discernible in this X-ray image, namely a discontinuity in the radial profile of X-ray brightness (but not in pressure) along each lobe  which is consistent with a cold front at the lobe's leading edge (\S \ref{xray_edge}). The distinct widening of both the remnant lobes near their outer edges would be consistent with this scenario. On the other hand, it is conceivable that a coherent magnetic field in the cold front has caused the sharp outer edges of the lobes A and B, as inferred from the
observed `double-scythe' like structure in A980 \citep{chibueze_2021Natur}. Clearly, a credible detection of such a structure is sensitively dependent on jet's orientation relative to the line-of-sight which is poorly constrained in the present case. But the non-detection of a compact radio component between A and B, and
their diffuse appearance with an ultra-steep radio spectrum, together constitute a strong hint that currently there are no jets feeding either of these lobes.

A likely process engendering the cold fronts detected in A980 is the outward pressure exerted due to the buoyancy-driven, near-sonically moving relic lobes \citep{2007PhR...443....1M}. However, an alternative to the cold-front scenario has been proposed for the case of NGC 507, the central galaxy of the Pisces cluster ($z$=0.01646). Its \textit{Chandra} image shows a prominent, sharp discontinuity in X-ray brightness (but only marginally in temperature) which is aligned with the outer edge of a radio lobe \citep{2004ApJ_kraft}. These authors have questioned the `cold-front' description of this system in view of (i) the sharpness and large magnitude of the brightness discontinuity, which is however not reflected in a temperature jump, and (ii) the lack of evidence for a radio lobe beyond the discontinuity, that could have maintained a pressure balance. They interpret the brightness discontinuity to arise substantially from an elemental abundance gradient across it (see, also, \citealt{2004ApJ_kempner}) and
find a higher elemental abundance in the cooler, low-entropy ICM within the cluster core, which has probably been entrained and transported outwards in the wake of the advancing radio lobes (see, e.g. \citealt{2003MNRAS_heinz, Paul_2020A&A} ). A potential caveat from the present perspective is that this argument based on abundance gradient works much more robustly for low-temperature ICM ($T<2$~keV) whose X-ray emissivity is high 
\citep{2004ApJ_kraft}.

\subsection{A radio mini-halo in formation?} \label{disc: mini_halo}

Our GMRT map of A980 at 325 MHz has revealed an extended diffuse radio source underlying the double radio source associated with the BCG (Fig.~\ref{fig:radio_high_opt}b). It is seen more clearly in the point source subtracted version of that map (Fig.~\ref{fig:A980_temerature} \& \S~\ref{sec:radio_anal}) where it is overlaid on the X-ray temperature map from our analysis of the \textit{Chandra} data (Fig.~\ref{fig:A980_temerature} \& \S~\ref{sec:xray_ktmap}). The diffuse radio source clearly overlaps the cool core, both having a size of $\sim100$~kpc. For this source with diffuse morphology, we estimate a flux density of $11.5\pm1.3$~mJy at 325~MHz which corresponds to a luminosity of $7.9 \pm 0.9 \times 10^{23}$ WHz$^{-1}$ and a radio emissivity of $1.1\pm0.2\times10^{-40}$~ergs$^{-1}$~cm$^{-3}$~Hz$^{-1}$ (see Appendix~\ref{apdx:emissivity}). With a size of $\sim100$~kpc (similar to the cool-core) and a characteristic higher radio emissivity compared to giant radio haloes \citep{Murgia_2009A&A}, the diffuse source detected around the BCG, can plausibly be classified as a radio mini-halo (e.g. \citealt{Giacintucci_2019ApJ}). 

A popular mechanism for such sources invokes re-acceleration of pre-existing low-energy electrons in the ICM, by turbulence induced by gas-sloshing \citep{2004ApJ_Fujita, 2013ApJ_zuhone}. In the present case, this could be caused by the postulated infall  of the BCG (\S \ref{Dying_radio_jets}) or possibly by the minor mergers hinted by the X-ray substructures roughly co-spatial with high-temperature patches (see \S~\ref{xray_properties} and \S~\ref{sec:xray_ktmap}) towards the north-east of the BCG. It may be noted that a similar hot patch has been reported for the ICM of the cool-core cluster RX~J1347.5-1145 \citep{Gitti_2004A&A}. Here, we may recall an alternative mechanism, according to which radio mini-haloes represent aggregate radio output from type-Ia supernovae occurring in the stellar population distributed across the cluster core \citep{2019MNRAS_omar}. If true, this alternative mechanism would obviate the need for the turbulent re-acceleration of relativistic particles, in order to create a radio mini-halo within the inner ICM. The mini-halo found here offers a rare opportunity to verify this alternative model. From Fig. \ref{fig:radio_high_opt}, it is seen that the prominent giant stellar halo of this BCG is elongated ($\sim 80$ kpc at position angle $\sim45$ deg). If improved radio imaging of this likely mini-halo reveals an extent and elongation matching that of the BCG stellar halo, which dominates the stellar content of the cluster core region, that would provide the much needed observational support to this (non-standard) supernova model of radio mini-halos.

\section{Conclusions}\label{conc}

We have investigated the sparsely studied galaxy cluster Abell 980 ($z = 0.1582$) by analysing its archival data from GMRT (150 and 325 MHz), EVLA (1.5 GHz) and {\it Chandra}. The ICM is found to have a quasi-relaxed appearance, with a cool core (ICM with $T\sim 4.2$~keV) of size $\sim 100$~kpc, surrounded by a large ($\sim 500$~kpc) hot gaseous halo ($T \sim 6.8$~keV) with two brightness discontinuities. The rich diversity of this cluster radio emission is found to include (i) two large diffuse sources of the ultra-steep spectrum (USS) located near the opposite extremities of the ICM, (ii) a bright double radio source ($\sim 55$ kpc) ) coincident with the BCG which lies close to the centre of the cluster's cool core, and (iii) a probable radio mini-halo of average size $\sim 110$ kpc associated with the BCG which itself possesses a huge ellipsoidal stellar halo of size ~ $\sim 80$ kpc. All four radio sources may have been created by the BCG at different stages and in different circumstances. The two large diffuse USS sources may be highly aged relics of the radio lobes produced by the BCG during its previous active phase  $\sim 260$~Myr ago and eventually propelled buoyantly towards the outskirts of the ICM, where they are found to create the two X-ray discontinuities in the ICM. The bright small double source with a central radio core is a probable outcome of a recent episode of activity in the BCG. Deeper radio imaging of the newly detected mini-halo presents a rare opportunity to verify the recent model in which mini-halos are explained as aggregate radio emission from Type I supernovae within the stellar population distributed across the cluster core. 

\begin{acknowledgements}: This research was funded by DST INSPIRE Faculty Scheme awarded to SP (code: IF-12/PH-44). S. Salunkhe would like to thank "Bhartratna JRD Tata Gunwant Sanshodhak Shishyavruti Yojna" for a doctoral fellowship.  G-K acknowledges a Senior Scientist fellowship from the Indian National Science Academy. S. Sonkamble acknowledges financial contribution from ASI-INAF n.2017-14-H.0 (PI A. Moretti). The authors gratefully acknowledge IUCAA, Pune, for partial financial support.
\end{acknowledgements}

\clearpage
\begin{appendix}
\section{The Radio data}

\subsection{Spectral index and error computation}\label{apdx:radio-spectrum}
GMRT 150 and 325~MHz images are made with uniform weight and appropriate uv-taper using the SPAM pipeline and are convoluted to a beam of size $33\arcsec\times22\arcsec$ and PA $60^\circ$ to make the spectral index map. The masking for this image was taken at the $3\sigma$ isophotes of the 325~MHz map. 

However, for the spectral index map between 1.5 GHz and 325~MHz, we made the maps using common uv-distance cut and uniform weights for both the frequencies and applied suitable uv-tapers in the uv-plane to get approximately the same resolution. Further, the maps were convoluted to a beam of $10'' \times 10''$ and were masked at the $3\sigma$ contour of the 1.5 GHz image.

\begin{figure}
    \begin{tikzpicture}
    \node(a){\includegraphics[width=0.46\textwidth]{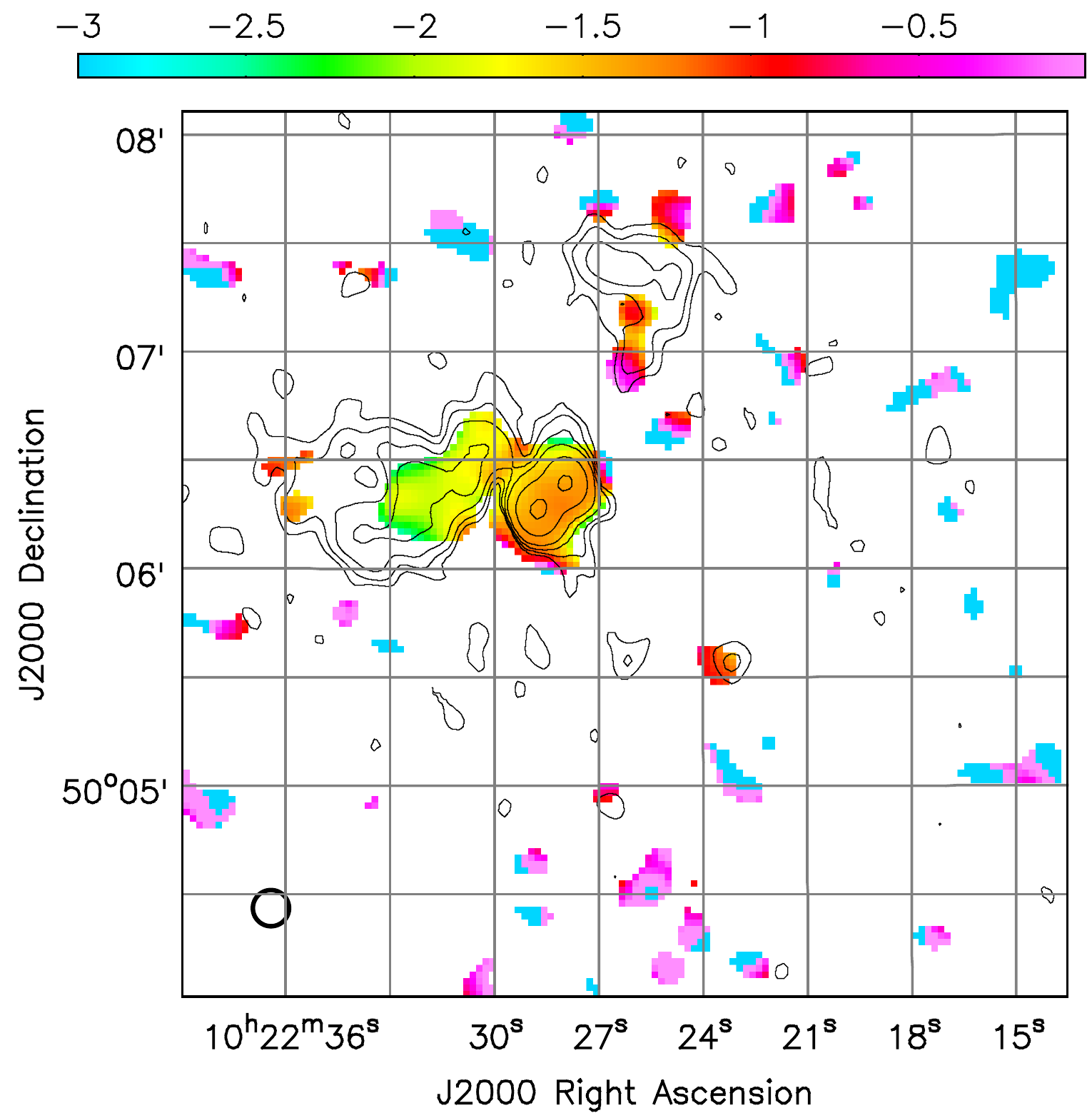}};
    \node at (a.south east)
    [
    anchor=center,
    xshift=-20mm,
    yshift=29mm
    ]
    {
    \includegraphics[width=0.18\textwidth]{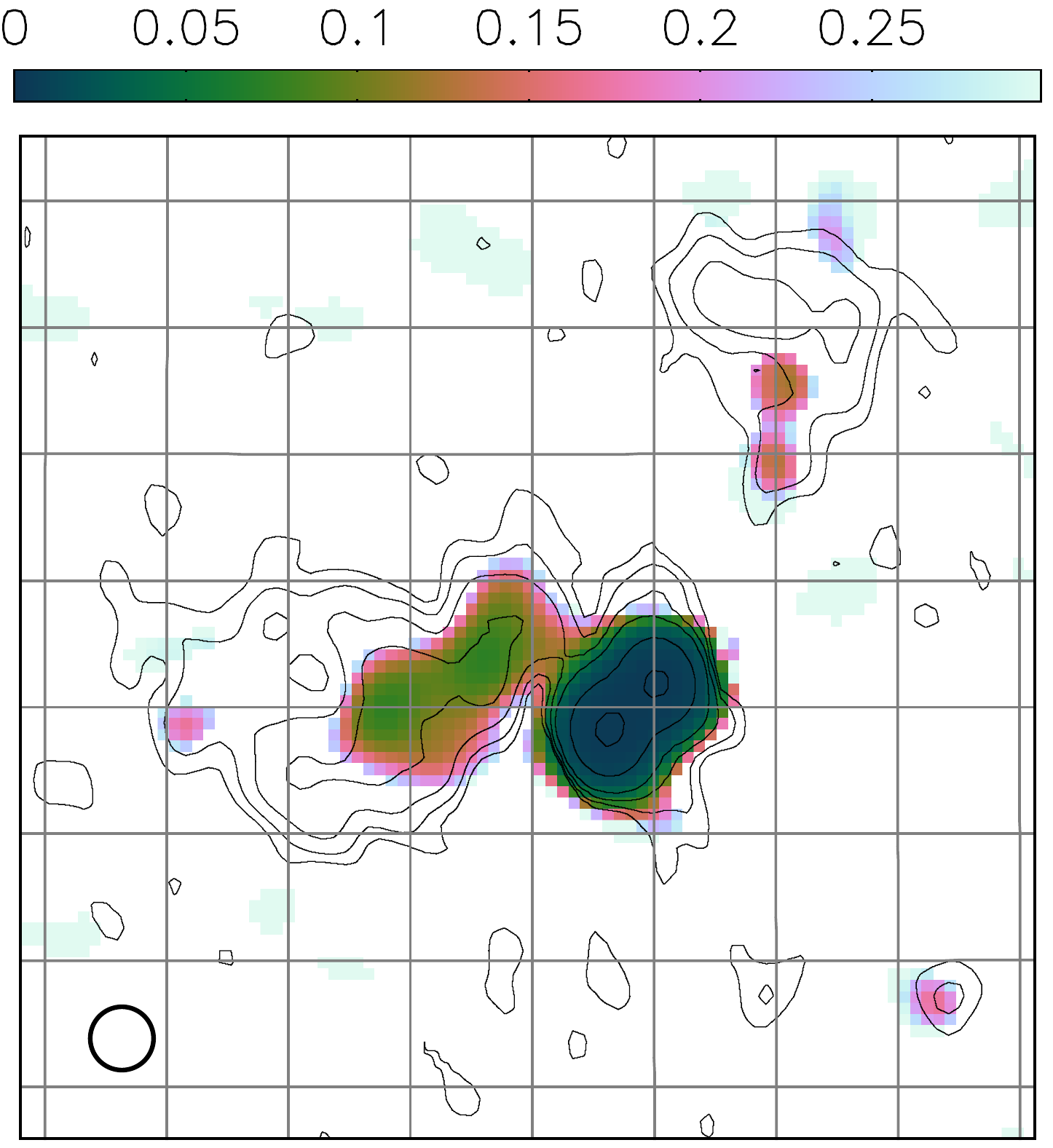}
    };
    \end{tikzpicture}
    \caption{spectral index and error maps for the cluster made using GMRT 325 MHz and EVLA L-band map} \label{spix} 
\end{figure}

The computation of the spectral index for both the cases were
is done using IMMATH task of CASA with the relation 
\begin{equation}\label{eq:Sp-Ind}
\alpha = \frac{\log(S_{\nu_2}/S_{\nu_1})}{\log(\nu_2/\nu_1)}
\end{equation}
Where $S_{\nu_\#}$ and $\nu_\#$ (with $\nu_1 > \nu_2$) are the values of flux density and the frequency of observation respectively. Spectral index maps between 150 and 325~MHz has been shown in Fig.~\ref{fig:spix-TGSS-GMRT} and between 325~MHz and 1.5~GHz in Fig.~\ref{spix}.

The spectral index error has been calculated with the same IMMATH task using the relation given below \citep{kim_trippe_2014_JKAS}
\begin{equation}\label{eq:SP-error}
\alpha_{err}(\alpha_{\nu_2,\nu_1})=\frac{1}{\log(\nu_2/\nu_1)}\times\left[\frac{\sigma_{\nu_1}^2}{I_{\nu_1}^2}+\frac{\sigma_{\nu_2}^2}{I_{\nu_2}^2}\right]^{\frac{1}{2}}
\end{equation}

With $I$ as the total intensity at respective frequencies at each of the pixels. Error maps (see the inset of Fig.~\ref{fig:spix-TGSS-GMRT}) shows a very reliable spectral map.

\subsection{Radio emissivity calculation}\label{apdx:emissivity}

Physical and morphological radio properties of the radio halo and mini-halo can be studied from their averaged brightness profiles. The emissivity of the radio halo or the mini-halo can be calculated if the central brightness and the e-folding radius are known which can be obtained by fitting the azimuthally averaged brightness profile with an exponential relation given below \citep{Murgia_2009A&A}.

\begin{equation}\label{halo_profile}
    I(r) = I_0 e^{-r/r_e}    
\end{equation}

Where central surface brightness ($I_0$) and e-folding radius ($r_e$) are two independent parameters. 

Analysis of the mini-halo becomes complicated because of the presence of the central radio galaxy. So, to separate the contribution of  this central source from the mini-halo, the point source subtracted image (see \S \ref{sec:radio_anal} for image analysis) is used for emissivity calculation. Concentric annuli of radius 3 arcsec to 30 arcsec are placed on a source spaced by 3 arcsec to get an azimuthally averaged brightness profile. Flux is measured within 3 sigma of noise level and also removed the contribution from source A. Obtained values of $I_0=31.5\pm 2.8\; \mu$Jy~arcsec$^{-1}$ and $r_e=41.3\pm2.8$~kpc after fitting are used in the following relation to calculate volume-averaged radio emissivity \citep{Murgia_2009A&A}.

\begin{equation}\label{emissivity}
    <J> \simeq 7.7 \times 10^{-41} (1+z)^{3+\alpha } \:. \frac{I_0}{r_e}
\end{equation}

Units of emissivity is erg~s$^{-1}\; \rm{cm}^{-3}\;\rm{Hz}^{-1}$ and used $I_0$ and $r_e$ are in units of $\mu$Jy~arcsec$^{-1}$ and kpc respectively. Emissivity for the mini-halo in A980 has been computed to be $1.1\pm0.2\times10^{-40}$ erg~s$^{-1}\: \rm{cm}^{-3} \: \rm{Hz}^{-1}$

\subsection{Magnetic field and the spectral age} \label{apdx: spectral_age}

If the break frequency ($\nu_{br}$)  and the magnetic field is known, one can obtain the age of the radio emitting electron source using the relation \citep{2004AA_jamrozy}. 
\begin{equation}
    \tau = 1.59 \times 10^3  \nu_{br}^{-0.5} \frac{B_{eq}^{0.5}}{B_{eq}^2+B_{CMB}^2} Myr.
\end{equation}
Where $B_{eq}$ is the equipartition field (in $\mu G$), $B_{CMB} = 3.25 (1+z)^2$ is the magnetic field equivalent to the
microwave background and $\nu_{br}$ is the break frequency (in GHz). Here we have assumed the break frequency at the lower end of our observation ($\nu_{br} = 150$ MHz). The equipartition magnetic field is calculated using the usual relation.

\begin{equation}
    B_{eq} = 7.91  \left[ \frac{1+k}{(1+z)^{\alpha - 3}}  \frac{S_0}{\nu_0^{\alpha} \theta_x \theta_y s} \frac{\nu_u^{\alpha+\frac{1}{2}} - \nu_l^{\alpha+\frac{1}{2}}}{\alpha+\frac{1}{2}}\right]^{\frac{2}{7}}
\end{equation}

where, $S_0$ is the flux density (in mJy) at the observing frequency ($\nu_0$), k is the ratio of the energy content in
relativistic protons to that of electrons, $\nu_l$ and $\nu_u$ are the lower and upper integration boundaries for the radio luminosity, $s$ is the path length through the source in the line of sight (in kpc), and $\theta_x$ and $\theta_y$ are source extents in two directions, measured in arcsec. For our case we have adopted k=1, $\nu_l =0.01$ GHz and $\nu_u = 100$ GHz. We have used $\theta_x$ and $\theta_y$ as the size of the sources `A' and `B' mentioned in \S \ref{sec:radio_properties} and the path length same as the average size of these sources.

\onecolumn
\section{The Xray data}
\vspace{-0.2cm}

\begin{table}[htp!]
\caption{Parameters of double $\beta$-model}
\centering
\small
\begin{tabular}{cccccccccc}\hline
Region  &$\beta$& $r_{c1}$ & $r_{c2}$ & ratio (R) & norm & const. & $\chi^{2}$/dof   \\
&&(arcmin)&(arcmin)&&($10^{-4}$)&($10^{-6}$) &     \\ \hline
0$\degr$ - 360$\degr$ & 0.65$\pm$0.05& 1.10$\pm$0.10   &0.03$\pm$0.01  &7.62$\pm$4.34 & 1.88$\pm$0.08 & 2.00$\pm$1.0 & 68.69/72 \\ 
\hline
\end{tabular}\\
 $r_{c1}$ - outer core radius, $r_{c2}$ - inner core radius, R - is the ratio between the two $\beta$ components at radius = 0
\label{tab:jump_SB}
\end{table}

\vspace{-0.3cm}
\begin{table}[!h]
\caption{Parameters of broken power-law density model}
\centering
\small
\begin{tabular}{ccccccccc}\hline
Region &$\alpha$1& $\alpha$2 & $r_{sh}$ & $n_0$ & Compression &$\chi^{2}$/dof   \\
&&&(arcmin)&($10^{-4}$)& (C) &     \\ \hline
10\degr - 100\degr & $0.62\pm0.06$   &$1.40\pm0.41$  &$1.30\pm0.06$&$0.50\pm0.13$ &$1.20\pm0.12$  &25.15/32 \\ 
175\degr - 243\degr & $0.26\pm0.17$   &$1.02\pm0.26$  &$1.32\pm0.10$&$0.65\pm0.16$ &$1.35\pm0.15$  &15.63/12 \\  \hline
\end{tabular}
\label{tab:jump_density}
\end{table}
\begin{multicols}{2}
\subsection{Extraction of global X-ray properties}\label{apdx:glob-xray}
The global X-ray properties of the cluster X-ray emission were determined by extracting a 0.5-8\,keV spectrum of the X-ray photons from within 150\arcsec\,(408\,kpc) circular region centred on the X-ray peak of the cluster. It is well-founded that the X-ray emission from the central giant elliptical galaxy in a cluster or group can significantly contaminate the estimate of ICM temperature near the cluster centre. To verify, how far the central BCG in our cluster contaminated our spectral fit, especially in the innermost region, we tried by excluding central 1, 2 and 3 arcsec and estimated the temperature and monitored the fit quality of the innermost region. We found that the fit with 2 arcsec exclusion provides the best-fit with a reduced Chi-square of 0.99. Also, the central 2 arcsec roughly contains 95 per cent of the encircled point-source energy \citep{Sansom_2006MNRAS}. Therefore, we finally excluded the central 2 arcsec region, so that the diffuse gas in the innermost region does not get contaminated by the central point source. A corresponding background spectrum was extracted from the normalized blank sky frame and appropriate responses were generated. The spectrum was binned such that every energy bin contains at least $\sim$20 counts and was then imported to {\tt XSPEC 12.9.1} for fitting adopting the $\chi^2$ statistics. The combined spectrum was fitted with an absorbed single temperature {\tt APEC} model with the Galactic absorption fixed at $N_{H}^{Gal} = 9.16\times10^{19}{\rm~cm^{-2}}$ \citep{2005A&A...440..775K}, letting the temperature, metallicity and normalization parameter to vary. The best fit resulted in the minimum $\chi^2=219.01$ for $210$ dof. 


\subsection{Surface Brightness Profile}\label{apdx:xray_SB}

We extracted azimuthal, blank sky background subtracted, exposure corrected surface brightness (SB) profile from a series concentric annuli (width = 2 arcsec) centred on the X-ray peak. Here we assumed spherical symmetry of the ICM. We used PROFFIT V 1.5 package\footnote{\color{blue} {http://www.isdc.unige.ch/\%7deckert/newsite/Proffit.html}} \citep{2011A&A...526A..79E} to fit the SB profile with a single $\beta$-model \citep{Cavaliere1976}, but, it did not fit well in the central region (see Fig.~\ref{fig:xray-fig} (b)). In this profile, one can clearly see the central brightness excess which is a typical characteristic of the cool-core system. We, therefore, added another $\beta$ component to model the central excess and the best-fitted parameters are shown in Table~\ref{tab:jump_SB}. The combined model (double $\beta$-model) is represented with the blue solid line in Fig.~\ref{fig:xray-fig} (b). This suggests that A980 is, indeed, a cool-core cluster.

\subsection{Edge characterization}\label{apdx:xray_edges}

We extracted the surface brightness profile up to a radius of 3\arcmin\, in the energy range of 0.5-3.0 keV along 10\degr - 100\degr and 175\degr - 243\degr  and fitted deprojected broken power-law density model within PROFFIT. The density model is defined as:
\begin{eqnarray}
	n(r) = \begin{cases} C\,n_{\rm {0}} \left(\frac{r}{r_{\rm front}}\right)^{-\alpha1}\,, & \mbox{if } r \le r_{\rm front} \\ n_{\rm {0}} \left(\frac{r}{r_{\rm front}}\right)^{-\alpha2}\,, & \mbox{if } r > r_{\rm front} \end{cases} \,,
\end{eqnarray}
where $n$ is the electron number density, $n_0$ is the density normalization, $C = n_{e_2}/n_{e_1}$ is the density compression factor, $\alpha$1 and $\alpha$2 are the power-law  indices, $r$ is the radius from the center of the sector and $r_{front}$ is the radius corresponding to the putative front. All the parameters of the model were kept free during the fit. The best-fitted profiles are shown in Fig.~\ref{fig:xray-fig}(c), while the best-fitted parameters are tabulated in Table~\ref{tab:jump_density}. From this profile, it is clear that there is a jump in the electron density at $\sim$1.3\arcmin\,(212 kpc) in both the regions. 

Further, we extract the spectrum from two regions (${\rm T_{1}}$ and ${\rm T_{2}}$) on either sides of the north-west edge and fit them with a single temperature {\tt APEC} model by keeping redshift fixed at 0.1582. The best fit temperature values for ${\rm T_{1}}$ and ${\rm T_{2}}$ are 5.77$\pm$0.48 keV~ and 10.90$\pm$3.33 keV, respectively. The ${\rm T_{1}}$ region contains hot, dense gas and a sharp boundary compared with ${\rm T_{2}}$. This indicates that the north-west edge is due to the presence of a cold front. We also extract the spectra on either side of the south-east edge and fit them in the same way, yielding best-fit temperatures of 5.53$\pm$0.52 keV and 12.22$\pm$3.87 keV, respectively. These derived temperature values seem to indicate that the south-east edge is also a cold front. 
We also obtained pressure for two cold fronts using $p = nkT$, where $n = n_e + n_H$ i.e total number density = electron density + hydrogen ion density, satisfy a relation $n_e = 1.2 n_H$. For pressure, we need electron density that was obtained from best fitted {\tt APEC} normalization following the method described in \cite{Kadam2019}. For north-west cold front the measured value of pressure across pre and post region are $5.38\pm0.72\times10^{-11}$ erg cm$^{-3}$, $4.49\pm0.40\times10^{-11}$ erg cm$^{-3}$, respectively. Whereas, for south-east cold fronts pressure values are $1.03\pm0.32\times10^{-11}$ erg cm$^{-3}$ , $1.04\pm0.10\times10^{-11}$ erg cm$^{-3}$. Therefore, these values suggested that the pressure is nearly continuous in pre and post cold-front regions, which agrees with the standard definition of a cold-front. However, the temperature measurement contains large uncertainties and therefore it is hard to arrive at a proper conclusion, a long exposure X-ray data would be needed to confirm these cold-fronts. 
\end{multicols}
\vspace{-0.5cm}
\end{appendix}

\vspace{-0.5cm}

\end{document}